\documentclass[12pt]{iopart}
\usepackage{epsfig}
\newcommand{\beq}{\begin{equation}}
\newcommand{\eeq}{\end{equation}}
\newcommand{\bea}{\begin{eqnarray}}
\newcommand{\eea}{\end{eqnarray}}
\newcommand{\bce}{\begin{center}}
\newcommand{\ece}{\end{center}}

\def\lsim{\mathrel{\rlap{\lower4pt\hbox{\hskip1pt$\sim$}}
    \raise1pt\hbox{$<$}}}         %less than or approx. symbol
\def\gsim{\mathrel{\rlap{\lower4pt\hbox{\hskip1pt$\sim$}}
    \raise1pt\hbox{$>$}}}         %greater than or approx. symbol

\begin{document}

\title{Heavy-Quark Diffusion, Flow and Recombination at RHIC} 

\author{Ralf Rapp\footnote[3]{email: rapp@comp.tamu.edu} and
Hendrik van Hees  
}

\address{Cyclotron Institute and Physics Department, Texas A\&M University, 
               College Station, Texas 77843-3366, U.S.A.}

\begin{abstract}
  We discuss recent developments in assessing heavy-quark interactions
  in the Quark-Gluon Plasma (QGP). While induced gluon radiation is
  expected to be the main energy-loss mechanism for fast-moving quarks,
  we focus on elastic scattering which prevails toward lower energies,
  evaluating both perturbative (gluon-exchange) and nonperturbative
  (resonance formation) interactions in the QGP. The latter are treated
  within an effective model for $D$- and $B$-meson resonances above
  $T_c$ as motivated by current QCD lattice calculations.  Pertinent
  diffusion and drag constants, following from a Fokker-Planck equation,
  are implemented into an expanding fireball model for Au-Au collisions
  at RHIC using relativistic Langevin simulations.  Heavy quarks are
  hadronized in a combined fragmentation and coalescence framework, and
  resulting electron-decay spectra are compared to recent RHIC data. A
  reasonable description of both nuclear suppression factors and
  elliptic flow up to momenta of $\sim$~5~GeV supports the notion of a
  strongly interacting QGP created at RHIC. Consequences and further
  tests of the proposed resonance interactions are discussed.
\end{abstract}

%Uncomment for PACS numbers title message
%\pacs{00.00, 20.00, 42.10}

% Uncomment for Submitted to journal title message
%\submitto{\JPG}

% Comment out if separate title page not required
%\maketitle

%%%%%%%%%%%%%%%%%%%%%%%%%%%%%%%%%%%%%%%%%%%%%%%%%%%%%%%%%%%%%%%%%%%%%%%%
\section{Introduction: The Virtue of Heavy Quarks in Heavy-Ion Collisions}
\label{sec_intro}
%%%%%%%%%%%%%%%%%%%%%%%%%%%%%%%%%%%%%%%%%%%%%%%%%%%%%%%%%%%%%%%%%%%%%%%%
The observed suppression of high-momentum particles in central Au-Au
collisions is one of the most exciting discoveries made at the
Relativistic Heavy-Ion Collider (RHIC) to date. It has been interpreted
as a large energy loss of energetic partons traveling through an almost
opaque partonic medium, the ``strongly-interacting Quark-Gluon Plasma''
(sQGP). The underlying microscopic mechanism has been attributed to
medium-induced radiation of gluons. However, recent data at RHIC,
showing a strong nuclear suppression (small
$R_{AA}$)~\cite{phenix-raa,star-raa} and elliptic flow (large
$v_2$)~\cite{phenix-v2} of "non-photonic" single electrons (attributed
to the semileptonic decay of charm and beauty hadrons) in semi-/central
Au-Au, have posed challenges to the radiative energy-loss picture within
perturbative QCD (pQCD). The importance of elastic interactions of heavy
quarks (HQs) in the QGP in this context has been pointed out in
Refs.~\cite{Hees04,Teaney04,Hees05,Djordjevic05,Goissaux06}.  Within
pQCD, radiative and elastic energy-loss mechanisms turn out to be
equally important in the currently accessible momentum range at RHIC,
and their combined effect reduces the discrepancy with the measured
electron $R_{AA}$ in central Au-Au (but does not eliminate
it)~\cite{Djordjevic05}. However, even when using upscaled transport
coefficients within pQCD energy-loss calculations~\cite{Armesto05}, the
maximal single-electron ($e^\pm$) $v_2$ is limited to 2-3\% in
semicentral Au-Au, while the experimental values reach up to 10\% around
$p_T^e$=2~GeV~\cite{phenix-v2}. Significantly larger theoretical $v_2$
values can only be obtained if the quarks interact strongly enough to
become part of the collectively expanding medium, implying that not only
energy-loss but also energy-gain processes (detailed balance) need to be
accounted for. This has recently been studied employing Langevin
simulations of HQs in an expanding QGP based on elastic
reinteractions~\cite{Hees04,Teaney04,Hees05,Goissaux06}. Since at RHIC
energies, secondary production of HQs is suppressed~\cite{LMW95}, the
latter are also valuable probes of the soft (collective) properties of
the putative sQGP, and can provide tests of hadronization mechanisms,
e.g., quark coalescence~\cite{Greco04}. HQ-momentum distributions
furthermore have important impact on secondary production (regeneration)
of heavy quarkonium states~\cite{GR02,GRB04,Thews06}.

In the present paper, we report on our studies~\cite{Hees04,Hees05} of
HQ properties in the sQGP and applications to RHIC, based on
non-perturbative interactions.  We will address pertinent uncertainties,
open problems and consequences of our findings thus far.

%%%%%%%%%%%%%%%%%%%%%%%%%%%%%%%%%%%%%%%%%%%%%%%%%%%%%%%%%%%%%%%%%%%%%%%%
\section{Heavy-Flavor Baseline Spectra at RHIC}
\label{sec_base}
%%%%%%%%%%%%%%%%%%%%%%%%%%%%%%%%%%%%%%%%%%%%%%%%%%%%%%%%%%%%%%%%%%%%%%%%
\begin{figure}[!b]
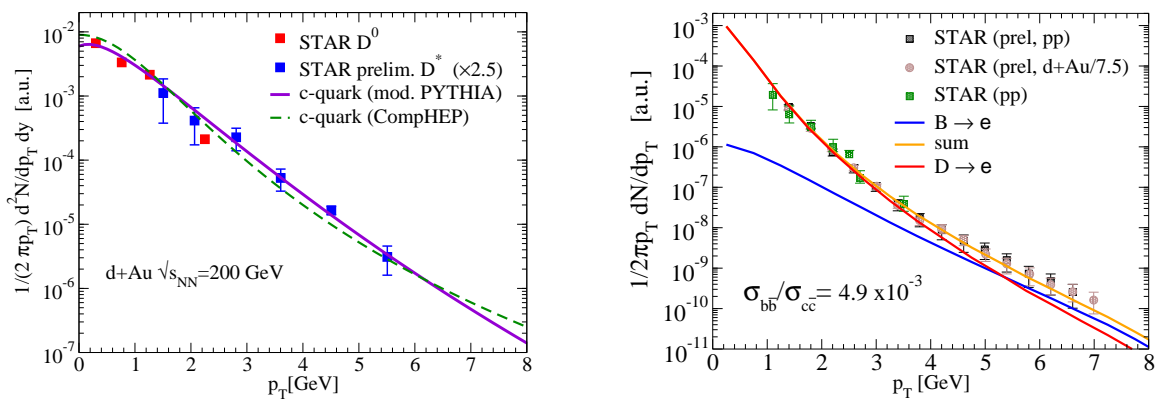

%\vspace{-0.8cm}
%\hspace{-0.5cm}
\begin{minipage}{7cm}
\epsfig{file=D-spectra-vs-c-Spectra.eps,width=7cm}
\end{minipage}
\hspace{1cm}
\begin{minipage}{7cm}
\epsfig{file=spectra-elect-charm-bottom-pp-new3.eps,width=7cm}
\end{minipage}
%\vspace{-0.5cm}
\caption{Left panel: $D$ and $D^*$ $p_T$-spectra
in $\sqrt{s_{NN}}$=200~GeV d-Au collisions~\cite{star-D};
right panel: composition of semileptonic electron-decay spectra
based on the charm spectra in the left panel in 
$p$-$p$ and d-Au at RHIC\cite{star-e-pp}.
}
\label{fig_base}
\end{figure}
The interpretation of semileptonic $e^\pm$-spectra in $A$-$A$ collisions
requires a reliable estimate of baseline spectra in $p$-$p$, in
particular their decomposition into contributions of charm and bottom
decays (bottom quarks are much more inert to changes in their
momentum spectra). Current pQCD predictions expect the transition from
charm to bottom at around $p_T^e\simeq$4-5GeV, albeit with significant
uncertainty~\cite{Cacciari05}. This rather early onset of a bottom
component constitutes part of the difficulty in explaining the large
suppression observed in the $e^\pm$-spectra in Au-Au collisions. In
Refs.~\cite{Hees05,Rapp05}, we have adopted a model-independent
procedure to determine $e^\pm$ baseline spectra, by first reproducing
available data on $D$ and $D^*$ spectra, evaluating their contribution
to the $e^\pm$ spectrum at low(er) $p_T$, and then adjusting the bottom
contribution to fill the high-$p_T$ part of the spectrum,
cf.~Fig.~\ref{fig_base}.  Our result confirms the crossing of the $c$
and $b$ contributions at about $p_T$$\simeq$5~GeV.

%%%%%%%%%%%%%%%%%%%%%%%%%%%%%%%%%%%%%%%%%%%%%%%%%%%%%%%%%%%%%%%%%%%%%%%%
\section{Elastic Heavy-Quark Scattering in the QGP}
\label{sec_elast}
\begin{figure}[!t]
\begin{minipage}{7.5cm}
\epsfig{file=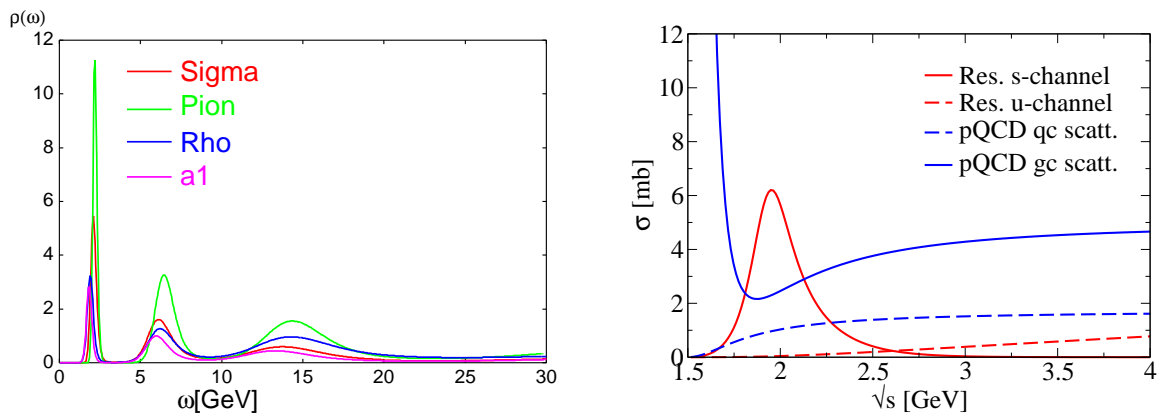,width=7.3cm}
\end{minipage}
\hspace{0.5cm}
\begin{minipage}{7cm}
\vspace{0.2cm}
\epsfig{file=charm-xsec.eps,width=7cm}
\end{minipage}
%\vspace{-0.5cm}
\caption{Left panel: mesonic spectral functions in the $q\bar q$ channel
for "strange" quarks~\cite{AH03}; note the approximate degeneracy of the 
chiral multiplets. Right panel: total cross sections for a $c$ quark 
scattering off partons within LO QCD (blue curves), and off antiquarks 
in the resonance model (red curves).}
\label{fig_xsec}
\end{figure}

\begin{figure}[!t]
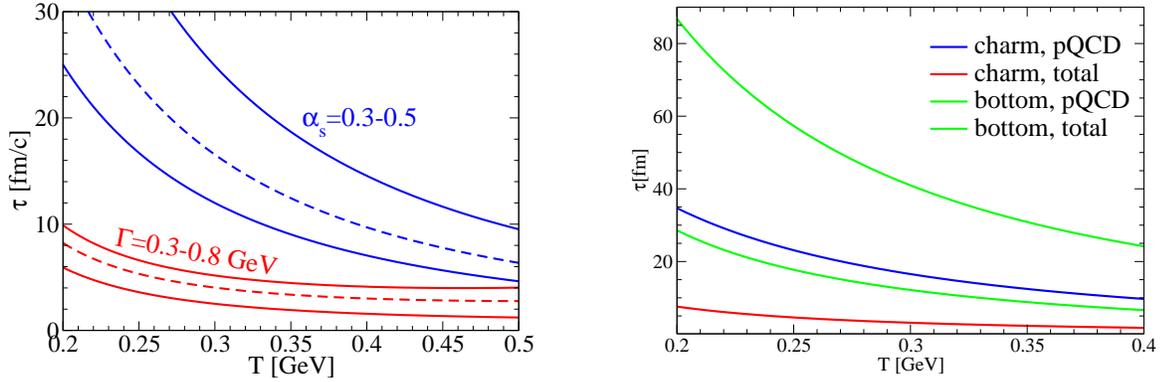

%\vspace{-0.8cm}
%\hspace{-0.5cm}
\begin{minipage}{7cm}
\epsfig{file=HQ-tau-vsT.eps,width=7cm}
\end{minipage}
\hspace{1cm}
\begin{minipage}{7cm}
\epsfig{file=tau-charm-bottom.eps,width=7cm}
\end{minipage}
%\vspace{-0.5cm}
\caption{Left panel: thermalization times of $c$ quarks in a QGP
using LO pQCD (upper band) and resonance interactions (lower band); 
right panel: $c$- and $b$-quark thermalization times for LO pQCD
interactions ($\alpha_s$=0.4, upper green and blue curve)
and when adding resonance interactions
($\Gamma_{\Phi}$=0.5~GeV, lower green and red curve).}
\label{fig_tau}
\end{figure}

%%%%%%%%%%%%%%%%%%%%%%%%%%%%%%%%%%%%%%%%%%%%%%%%%%%%%%%%%%%%%%%%%%%%%%%%
We evaluate (elastic) interactions of HQs in the QGP using Brownian
motion of a heavy particle in a thermal background of light partons, as
encoded in a Fokker-Planck
equation~\cite{Svet88,Mustafa98,Hees04,Teaney04},
\begin{equation}
\frac{\partial f}{\partial t} = \gamma \frac{\partial (p f)}{\partial p}
+ D \frac{\partial^2 f}{\partial p^2} \ , 
 \end{equation}
 for the HQ distribution function, $f$. $\gamma$=$\tau_Q^{-1}$ and $D_p$
 are the corresponding drag and (momentum) diffusion constants which
 determine the approach to equilibrium and satisfy the Einstein
 relation, $T=D_p/\gamma M_Q$. They are typically calculated from
 elastic 2$\leftrightarrow$2 scattering processes, $p +Q \to p+Q$
 ($p$=$q$, $\bar q$, $g$; $Q$=$c$, $b$). In leading-order (LO) pQCD these are
 dominated by $t$-channel gluon exchange (regulated by a Debye mass).
 E.g., at a temperature $T$=300~MeV, and for a strong coupling constant
 $\alpha_s$=0.4, the thermal relaxation time for charm is
 $\tau_c$$\simeq$15~fm/$c$, well above expected QGP lifetimes of
 $\tau_{QGP}$$\lsim$5~fm/$c$ at RHIC.  On the other hand, lattice QCD
 suggests that hadronic resonance (or bound) states in both $Q$-$\bar
 Q$ and $q$-$\bar q$ channels might survive up to
 temperatures of $\sim$2$T_c$~\cite{AH03,KL03}, cf.~left panel of
 Fig.~\ref{fig_xsec}.
In Ref.~\cite{Hees04} we therefore suggested that $D$- and $B$-meson
resonances in the $Q$-$\bar q$ channel could accelerate the thermal
relaxation of HQs. Based on an effective Lagrangian for the $\bar
q$-$Q$-$\Phi$ interaction ($\Phi$=$D$, $B$),
\begin{equation}
{\cal L} = Q~\frac{1+\not{\!v}}{2}~\Phi~\Gamma~\bar q + {\rm h.c}  
\label{lag}
\end{equation}
we evaluated elastic $Q+\bar q \to Q+\bar q$ scattering amplitudes
via $\Phi$ exchange in $s$-, $t$ and $u$-channel. Assuming the
existence of one $\Phi$ state (e.g., a pseudoscalar $J^P$=$0^-$), and 
a minimal degeneracy following from chiral and HQ symmetries (with Dirac 
matrices $\Gamma$=1, $\gamma_5$, $\gamma^\mu$, $\gamma_5\gamma^\mu$
in Eq.~(\ref{lag})), the $\tau_Q$'s
are reduced by a factor $\sim$3 over pQCD scattering
(cf.~Fig.~\ref{fig_tau}), with moderate sensitivity to the unknown 
coupling at the effective $c$-$\bar q$-$D$ ($b$-$\bar q$-$B$) vertex.
While the total cross sections for LO pQCD and resonance scattering are 
not very different in magnitude (right panel in Fig.~\ref{fig_xsec}), 
the angular distributions are: forward-dominated pQCD scattering is much 
less efficient in isotropizing the momentum distributions compared to
the isotropic resonance scattering, since the (angular-averaged) 
"transport cross section" carries a weight of $1-\cos \theta$ 
($\theta$: scattering angle in the center of mass).  

%%%%%%%%%%%%%%%%%%%%%%%%%%%%%%%%%%%%%%%%%%%%%%%%%%%%%%%%%%%%%%%%%%%%%%%%
\section{Heavy-Quark and Single-Electron Spectra at RHIC}
\label{sec_rhic}
%%%%%%%%%%%%%%%%%%%%%%%%%%%%%%%%%%%%%%%%%%%%%%%%%%%%%%%%%%%%%%%%%%%%%%%%
\begin{figure}[!b]
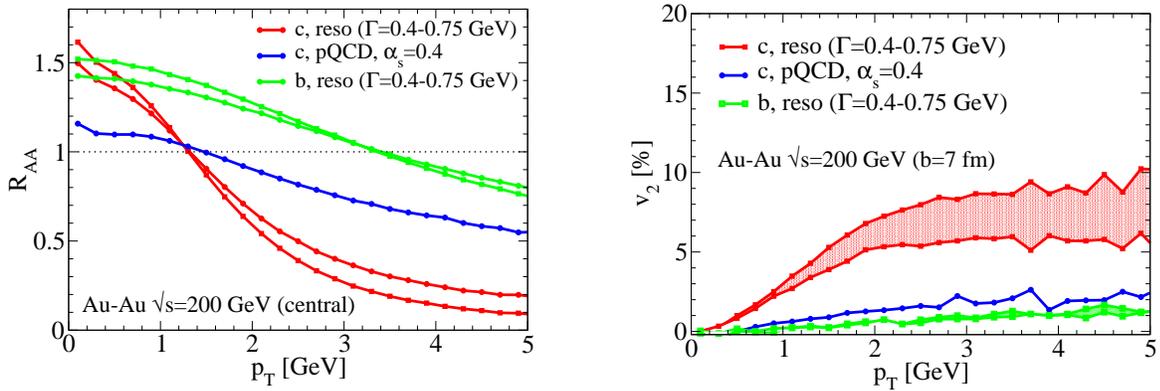

%\vspace{-0.8cm}
%\hspace{-0.5cm}
\begin{minipage}{7cm}
\epsfig{file=quark-RAA-central.eps,width=7cm}
\end{minipage}
\hspace{1cm}
\begin{minipage}{7cm}
\epsfig{file=quark-minbias-v2.eps,width=7cm}
\end{minipage}
%\vspace{-0.5cm}
\caption{HQ nuclear suppression factor (left panel) and elliptic flow  
(right panel) following from relativistic Langevin simulations for
an expanding QGP in Au-Au at RHIC; red (green) lines: $c$ ($b$) quarks
using resonance+pQCD interactions; blue lines: $c$ quarks with LO
pQCD scattering only.}
\label{fig_quark}
\end{figure}
Using the initial $c$- and $b$-quark spectra as discussed in
Sec.~\ref{sec_base}, the diffusion and drag coefficients as evaluated in
Sec.~\ref{sec_elast} have been implemented into a relativistic Langevin
simulation for an expanding QGP fireball at RHIC~\cite{Hees05}. As
expected, resonance interactions have a much larger effect on the
nuclear suppression factor ($R_{AA}$) and elliptic flow ($v_2$) of $c$
quarks in semi-/central Au-Au collisions than pQCD interactions
(cf.~Fig.~\ref{fig_quark}), while $b$ quarks are less affected. The
leveling-off of the $c$-quark elliptic flow characterizes the transition
from a quasi-thermal to a kinetic regime.
 
\begin{figure}[!t]
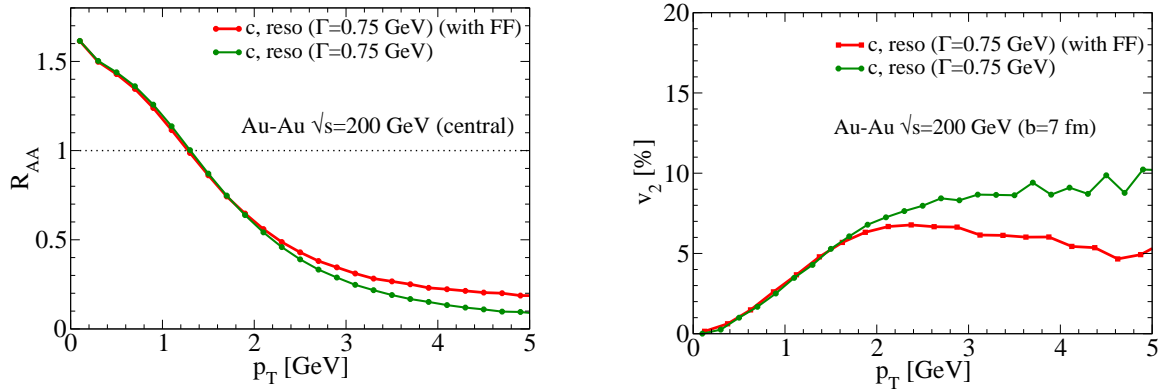

\begin{minipage}{7cm}
\epsfig{file=quark-RAA-central-FF1.0-vs-renorm.eps,width=7cm}
\end{minipage}
\hspace{1cm}
\begin{minipage}{7cm}
\epsfig{file=quark-v2-semicentral-FF1.0-vs-renorm.eps,width=7cm}
\end{minipage}
\caption{Comparison of the $c$-quark $R_{AA}$ and $v_2$ following
from Langevin simulations at RHIC when using effective $\bar q$-$c$-$D$
vertices with renormalization (green curves) or monopole form factors 
with cutoff $\Lambda$=1~GeV (red curves).}
\label{fig_eq}
\end{figure}

Fig.~\ref{fig_eq} illustrates the sensitivity of the Langevin results to
the regularization procedure applied to the $Q$-$\bar q$
loops~\cite{Hees04}: at quark momenta above 3~GeV a renormalized
interaction (point-like vertices) leads to somewhat stronger effects
compared to finite-size vertices (form factor with cutoff
$\Lambda$=1~GeV and identical resonance widths).

\begin{figure}
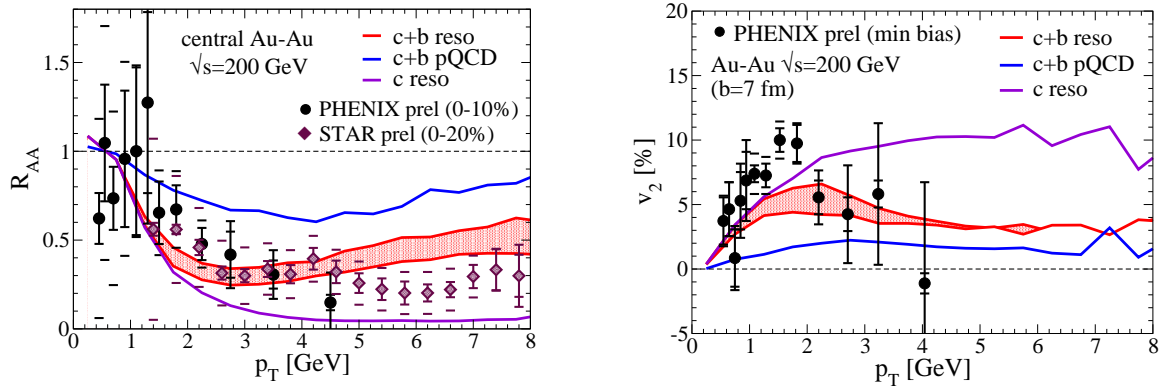

\begin{minipage}{7cm}
\epsfig{file=raa_e_cent_fragonly-qm05data.eps,width=7cm}
\end{minipage}
\hspace{1cm}
\begin{minipage}{7cm}
\epsfig{file=v2_e_MB_delta_frag_only-qm05data.eps,width=7cm}
\end{minipage}
\caption{Nuclear suppression factor (left panel) and elliptic flow
(right panel) for non-photonic single-electron spectra in semi-/central
Au-Au collisions at RHIC. Data~\cite{phenix-raa,star-raa,phenix-v2} are 
compared to theory predictions~\cite{Hees05}
using Langevin simulations with elastic $c$- and $b$-quark interactions 
in an expanding QGP fireball and $\delta$-fragmentation at $T_c$; 
red band: 
LO pQCD + resonance interactions (with $\Gamma_\Phi$=0.4-0.75~GeV), 
blue line: LO-pQCD only, purple line: $c$ quarks only for LO pQCD
and resonances.}
\label{fig_frag}
\end{figure}

\begin{figure}
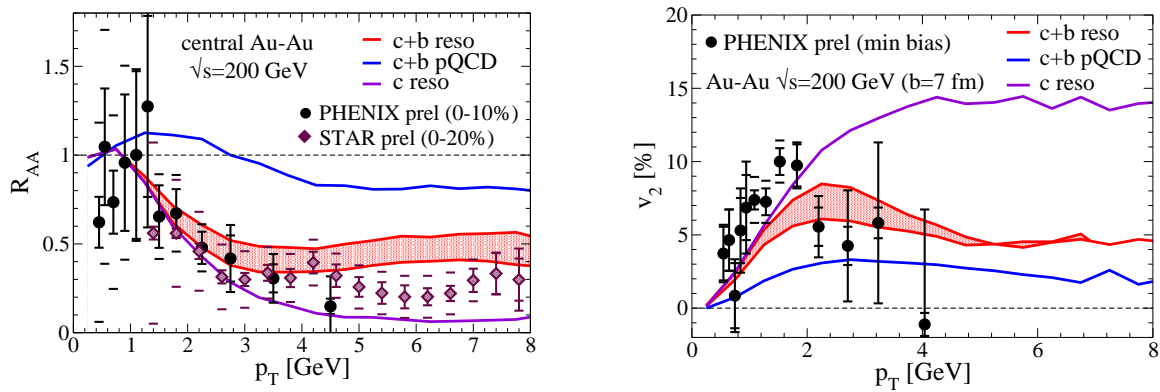

\begin{minipage}{7cm}
\epsfig{file=raa_e_cent-therm.eps,width=7cm}
\end{minipage}
\hspace{1cm}
\begin{minipage}{7cm}
\epsfig{file=v2_e_MB_therm-av-qm05data.eps,width=7cm}
\end{minipage}
\caption{Same as Fig.~\ref{fig_frag} but including heavy-light
quark-coalescence at hadronization.}
\label{fig_coal}
\end{figure}

To compare to $e^\pm$ observables, the HQs have been hadronized and
decayed semileptonically~\cite{Greco04,Hees05}. If the hadronization is 
performed by using $\delta$-function fragmentation only, the resulting 
$e^\pm$ $R_{AA}$ and
$v_2$ do not describe the data very well, cf.~Fig.~\ref{fig_frag}.  
The situation improves when allowing for coalescence of the HQs
with light quarks~\cite{Greco04,Hees05} at $T_c$ which
preferentially occurs at lower momenta (the remaining 
HQs are again $\delta$-function fragmented).  
Consequently, we find an increase in \emph{both} $R_{AA}$ and $v_2$
mostly for momenta up to $p_T$$\simeq$4-5~GeV, leading to
fair agreement with experiment, cf.~Fig.\ref{fig_coal}.
At higher $p_T$, contributions from radiative energy-loss 
(which have been neglected here) are expected to become important.  
Note that at high $p_T$ the data for $R_{AA}$ stay small, while the $v_2$
tends to (much) reduced values, which is suggestive for substantial energy 
loss with small collectivity. Such a feature is indeed consistent with
perturbative energy-loss mechanisms. We also recall that the Langevin
results in Figs.~\ref{fig_frag} and \ref{fig_coal} presumably
overestimate the suppression and $v_2$ at high $p_T$ somewhat, due to
the assumptions of (i) point-like vertices (cf. Fig.~\ref{fig_eq}), and
(ii) a thermal background medium even at high $p_T$ (while from light
hadron spectra of RHIC one infers that full thermalization for light
partons reaches to about $p_T$$\simeq$1~GeV).

%%%%%%%%%%%%%%%%%%%%%%%%%%%%%%%%%%%%%%%%%%%%%%%%%%%%%%%%%%%%%%%%%%%%%%%%
\section{Discussion and Further Tests}
\label{sec_quarkonia}
%%%%%%%%%%%%%%%%%%%%%%%%%%%%%%%%%%%%%%%%%%%%%%%%%%%%%%%%%%%%%%%%%%%%%%%%
The small $R_{AA}$ and the large $v_2$ in the $e^\pm$ data require
strong interactions of heavy quarks in the QGP, (well) beyond radiative
energy-loss. While elastic pQCD scattering reduces the discrepancy in
the $R_{AA}$, a sufficient collectivity to explain the $v_2$ can only be
obtained for (unrealistically!?) large coupling constants~\cite{Teaney04}.
Resonance scattering is a promising candidate for a microscopic
explanation of the apparently large diffusion constant (for an estimate
of energy loss due to 3$\leftrightarrow$3 scattering,
cf.~Ref.~\cite{LK06}).

An important feature that distinguishes resonance contributions from
pQCD relates to the chemical composition of the QGP. While the
resonances as implemented here require the presence of anti-/quarks
(here we have assumed QGP in chemical equilibrium for 2+1 flavors,
amounting to $\sim$25 anti-/quark and 16 gluon degrees of freedom), pQCD
approaches typically assume a gluon plasma by saturating the bound 
on the produced entropy in central Au-Au at RHIC with 
$\mathrm{d}N_g/\mathrm{d}y$=1000; this
maximizes the color charge and thus the interaction strength in both
radiative (gluon emission) and elastic (gluon exchange) channels. Thus,
if RHIC indeed produces a gluon plasma, resonance scattering could be
severely suppressed (much less is known about the presence of colored
composites~\cite{SZ04}). 
However, if quark-pair production occurs early in the collision
resulting in an approximately chemically equilibrated 
QGP~\cite{Lappi06}, more than half of the
partons are anti-/quarks and the pQCD transport coefficient is
appreciably reduced, increasing the importance of non-perturbative
effects.

While the existence of resonances needs to be scrutinized theoretically
(e.g., by lQCD calculations of heavy-light correlators, or effective
models based on lQCD potentials~\cite{SZ04,Mannarelli:2005}), one should 
also search for experimental means of discrimination. The relative 
quark content of the colliding nuclear system can be increased 
by lowering the center-of-mass energy, while antiquarks (and possibly
gluons) become suppressed. Thus, one would expect
stronger rescattering of $\bar c$ quarks compared to $c$ quarks,
resulting in larger $v_2$ and smaller $R_{AA}$ for $D^-$ and $\overline{D}^0$
relative to $D^+$ and $D^0$-mesons. Whether such an effect is detectable
at lower RHIC energies, or whether one has to await the (higher 
luminosities at the fixed-target) Compressed
Baryonic Matter (CBM) experiment at the future GSI facility, needs to be
checked in quantitative studies. We also note that
diquark-like $c$-$q$ interactions could diminish the differences between
$D$ and $\overline{D}$ mesons (or even charm-exchange reactions in the
hadronic phase~\cite{Cass06}).

Finally, it is important to keep in mind the consistency of open and
hidden charm. As has been emphasized by several
authors~\cite{pbm01,GR02,GRB04,Thews06}, secondary production
(regeneration) of charmonia is much facilitated by a softening
(thermalization) of the $c$- and $\bar c$-quark momentum distributions
(for bottom/onia, this interplay appears to be less pronounced, even at
the LHC~\cite{Grandchamp05}).

%%%%%%%%%%%%%%%%%%%%%%%%%%%%%%%%%%%%%%%%%%%%%%%%%%%%%%%%%%%%%%%%%%%%%%%%
\section{Conclusions}
\label{sec_concl}
%%%%%%%%%%%%%%%%%%%%%%%%%%%%%%%%%%%%%%%%%%%%%%%%%%%%%%%%%%%%%%%%%%%%%%%%

Heavy-quark probes at RHIC have posed new challenges to the theoretical
understanding of parton-energy loss at high momenta and (the approach
to) equilibration at moderate and low momenta. Currently, the severeness
of this problem hinges on the interpretation of (semileptonic-decay)
electron spectra, and in particular their relative decomposition into
charm- and bottom-decay contributions (obviously, explicit $D$-meson 
measurements would remove this ambiguity). If their crossing in baseline
$p$-$p$ spectra occurs around $p_T$$\simeq$5~GeV, the combination of
radiative and elastic pQCD interactions of heavy quarks is apparently
too weak to generate a suppression and collectivity consistent with the
observed spectra in semi-/central Au-Au collisions.  We have presented a
scenario based on resonance exchange to introduce nonperturbative
elastic interactions of heavy quarks in the sQGP at moderate
temperatures (relevant for RHIC).  When treated within a Brownian motion
scheme for an expanding QGP fireball, a fair description of the
single-$e^\pm$ spectra up to $p_T$$\simeq$4-5~GeV emerges (especially if
hadronization is supplemented with heavy-light-quark coalescence); at
higher $p_T$, induced gluon radiation (not included here) is expected to
take over and become the main source of heavy-quark energy loss. A
consistent implementation of the latter in the Fokker-Planck approach,
which requires the integration of coherence effects, is not an easy
task. We have also emphasized the importance of the chemical composition
of the (early) QGP which affects the diffusion constants in the
perturbative and nonperturbative sectors (as evaluated here) in a rather
opposite way and may thus help to disentangle the two. An excitation
function could provide a valuable handle to explicitly vary the
quark-to-gluon ratio to further illuminate the nature of heavy-quark
diffusion in the QGP.

\vspace{0.5cm}

\noindent
{\bf Acknowledgment} \\
One of us (RR) would like to thank the conference organizers for the
invitation to a very informative meeting.  This work was supported in
part by a U.S.~National Science Foundation CAREER award under grant no.
PHY-0449489.

\end{document}